\documentclass[aps,pre,preprint,superscriptaddress]{revtex4-1} 

\usepackage{graphicx}
\usepackage{amsmath}
\usepackage{amssymb}
\usepackage{mathrsfs} 
\usepackage{color} 

\providecommand\bcdot{\boldsymbol{\cdot}}

\providecommand\bx{\mathbf{x}}

\providecommand\be{\mathbf{\hat{e}}}

\newcommand{\pd}[2]{\frac{\partial #1}{\partial #2}}

\newcommand{\ub}[1]{^{({#1})}}

\newcommand{\TheTitle}{Bloch waves in an arbitrary two-dimensional lattice of subwavelength Dirichlet scatterers}

\begin{document}

\begin{abstract}
We study waves governed by the planar Helmholtz equation, propagating in an infinite lattice of subwavelength Dirichlet scatterers, the periodicity being comparable to the wavelength. Applying the method of matched asymptotic expansions, the scatterers are effectively replaced by asymptotic point constraints. The resulting coarse-grained Bloch-wave dispersion problem is solved by a generalised Fourier series, whose singular asymptotics in the vicinities of scatterers yield the dispersion relation governing modes that are strongly perturbed from plane-wave solutions existing in the absence of the scatterers; there are also empty-lattice waves that are only weakly perturbed. Characterising the latter is useful in interpreting and potentially designing the dispersion diagrams of such lattices. The method presented, that simplifies and expands on Krynkin \& McIver [Waves Random Complex, \textbf{19} 347 2009], could be applied in the future to study more sophisticated designs entailing resonant subwavelength elements distributed over a lattice with periodicity on the order of the operating wavelength. 
\end{abstract}

\title{{\TheTitle}}

\author{O.~Schnitzer}
\affiliation{Department of Mathematics, Imperial College London, London SW7 2AZ, UK}
\author{R.~V.~Craster}
\affiliation{Department of Mathematics, Imperial College London, London SW7 2AZ, UK}

\maketitle



\section{Introduction}
There is immense current interest in wave phenomena in artificial periodic media. In subwavelength metamaterials (SWM), that are constructed using tiny resonant elements, the periodicity is small compared with the operating wavelength \cite{Smith:04}. In a sense, SWM mimic natural materials whose macroscopic properties are endowed by an underlying atomic structure. Ingenious designs --- introduced in electromagnetics but since adapted to acoustics, elasticity and seismology --- have enabled however quite unnatural effective properties and capabilities, including a negative refractive index, cloaking, and subwavelength imaging. Photonic (similarly platonic, phononic, \textit{etc.}) crystals constitute a separate class of artificial materials, in which operating wavelengths are typically on the order of the periodicity of the microstructure \cite{Sakoda:Book}. Here, wave manipulation is enabled by the surprisingly coherent outcome of multiple scattering events, mimicking the way electron waves are sculptured in solid-state crystals. Studies of photonic crystals were initially focused on the existence of complete photonic band gaps \cite{Yablonovitch:87}. Nowadays, however, artificial crystals broadly interpreted are in the spotlight as a lossless alternative to SWM, with the plethora of phenomena demonstrated including slow light \cite{Baba:08}, self collimation \cite{Kosaka:99}, dynamic anisotropy \cite{Antonakakis:14}, defect and interface modes \cite{Joannopoulos:Book}, all-angle-negative refraction \cite{luo:02}, subwavelength imaging \cite{Luo:03}, cloaking \cite{Ergin:10}, topologically protected edge states \cite{Lu:14}, and unidirectional propagation \cite{Mcphedran:15,colquitt:15} amongst others.

Artificial materials made out of subwavelength particles, inclusions, or microstructured resonators, periodically distributed with spacing on the order of the operating wavelength, can be thought as being somewhere in between SWM and photonic crystals. The smallness of the scattering elements in this setup suggests a relatively weak modulation of waves propagating through the crystal. When this is indeed the case, the effect of the lattice can be captured by a perturbation approach \cite{Mciver:07} in the spirit of the ``empty-lattice approximation'' popularised in solid-state physics \cite{Kittel:Book}. There are scenarios, however, where the modulation cannot be reasonably regarded as small, regardless of the scatterer dimensions. An elementary example is the propagation of flexural waves in a periodically pinned plate, where the time-harmonic displacement field is governed by the two-dimensional Biharmonic reduced-wave equation and a zero-displacement ``Dirichlet'' condition on pinned boundaries. Since the Green function of the latter equation is regular, a valid leading-order approximation is obtained by simply ignoring the finite size of the scatterers, treating these as Dirichlet point constraints. The resulting problem offers a convenient framework for detailed mathematical studies of Bloch-wave dispersion surfaces and scattering from finite and semi-infinite arrays \cite{Evans:07,Guo:11,Smith:14,Haslinger:15}, and for demonstrating the method of ``High-Frequency Homogenisation'' \cite{Antonakakis:13,Makwana:16}.

A related scenario is the propagation of planar electromagnetic waves through a doubly periodic array of perfectly conducting cylinders in transverse-magnetic polarisation, where the time-harmonic electric-field component along the cylindrical axes satisfies the Helmholtz equation and vanishes on the cylindrical boundaries \cite{Nicorovici:95}. The acoustic analogue of the latter problem entails soft cylinders on which the velocity potential vanishes. Mathematically, the singular nature of the Helmholtz Green function implies that, contrary to the Biharmonic case, one cannot prescribe the value of the wave field at a point. Accordingly, the small-scatterer limit needs to be handled with care. One approach is to reduce an exact multipole-expansion solution in the latter limit \cite{Martin:10}. A more general and systematic approach is to employ the method of matched asymptotic expansions \cite{Hinch:91}, where ``inner'' expansions valid in the vicinities of the scatterers are matched with an ``outer'' expansion valid away from the scatterers. The pertinent essentials of the approach are didactically outlined in Refs.~\cite{Crighton:12} and \cite{Martin:06} in the context of low-frequency scattering from isolated subwavelength Dirichlet scatterers, and in the context of singularly perturbed eigenvalues of the Laplace operator in Ref.~\cite{Ward:93:strong}. 

A matched asymptotics solution of the Bloch-wave dispersion problem for a doubly periodic array of small cylindrical Dirichlet scatterers was given by Krynkin and McIver \cite{Krynkin:09}. 
In that work, the analysis of the inner region and the matching procedure follow closely the approach in Ref.~\cite{Crighton:12}, generalised to a cylinder of arbitrary cross-sectional shape. The outer solution is sought from the start as a quasi-periodic Green function, namely an infinite double sum of Hankel functions whose convergence is accelerated using the addition theorem and highly developed techniques for manipulation of lattice sums. The main result is an asymptotic dispersion relation, for which several accurate and semi-accurate methods of solution are suggested. A key finding is that deviations from  empty-lattice eigenvalues, identified there as poles of lattice sums, can be either appreciable or negligible, i.e., logarithmically or algebraically small with respect to the inclusion size. 

In this paper, we revisit the doubly periodic Helmholtz--Dirichlet problem. We aim to streamline the approach of Krynkin and McIver \cite{Krynkin:09} thereby allowing the methodology to be more readily applied, and also to generalise to lattices with elementary unit cells occupied by multiple scatterers (e.g.~the honeycomb lattice that is of much current interest due to its connection to graphene and isolated Dirac points). In our approach, those empty-lattice Bloch waves that are weakly perturbed are systematically identified \emph{a priori} by considering the existence (and dimension of the space) of empty-lattice waves that vanish at the positions of all scatterers. In addition to simplifying the analysis, this approach provides insight when interpreting the resulting dispersion surfaces, and potentially in the design of tailored media. {We also take a different matching route from} that of Krynkin and McIver \cite{Krynkin:09}. They adopt an infinite asymptotic expansion in inverse logarithmic powers, and following \cite{Crighton:12}, identify, in the course of the analysis, a Poincar\'e perturbation parameter, corresponding to a summation of the infinite logarithmic series. This parameter, however, is taken as a logarithmic perturbation to a leading $O(1)$ term, the asymptotic separation being comparable to that intended to be avoided in the first place. In contrast, we adopt the notion of treating logarithmic terms on par with $O(1)$ terms \cite{Crighton:73,Keller:76}, which  
{renders} the matching procedure to be more straightforward. 
{Perhaps the} key difference is that we bypass the machinery of addition theorems and lattice-sum acceleration associated with multipoles. Rather, in the case of strongly perturbed modes, we represent the outer-region solution in terms of a generalised Fourier series. The dispersion relation is then derived by matching the singular asymptotics of the latter in the vicinities of the scatterers with the respective inner solutions. The present analysis {thereby} corrects the Fourier-series example of Refs.~\cite{Antonakakis:13} and \cite{Makwana:16} for the singular Helmholtz case; the Fourier series solutions there were prescribed to vanish at the scatterers' positions, whereas those series actually diverge there. 

In \S\S\ref{ssec:form} we formulate the Bloch-wave dispersion problem for an arbitrary two-dimensional Bravais lattice of small Dirichlet scatterers. In \S\S\ref{ssec:matching} we outline the asymptotic procedure for replacing the finite scatterers with point-singularity constraints, leading to the coarse-grained eigenvalue problem of \S\S\ref{ssec:effective}. Weakly and strongly perturbed modes are analysed in \S\S\ref{ssec:weak} and \S\S\ref{ssec:strong}, respectively, and dispersion diagrams for square and hexagonal lattices are shown and discussed in \S\S\ref{ssec:curves}. The approach is generalised in  \S\ref{sec:multiple}  to lattices with multiply occupied elementary cells, and the dispersion diagram for a honeycomb lattice is shown and interpreted. In \S\ref{sec:scat} we briefly consider the applicability of the asymptotic procedure to solving scattering problems involving a finite collection of scatterers, and employ the resulting scheme towards demonstrating the strong dynamic anisotropy suggested by the infinite-lattice dispersion surfaces. Lastly, in \S\ref{sec:disc} we give concluding remarks and discuss further research directions.

\section{Small Dirichlet scatterers at Bravais lattice points} \label{sec:main_analysis}
\subsection{Problem formulation} \label{ssec:form}
Consider the dimensionless Helmholtz equation in two dimensions,
\begin{equation}\label{helm}
\nabla^2u+\omega^2u=0.
\end{equation}
Here $u(\bx)$ is an arbitrarily normalised field, $\bx$ being the position vector normalised by a characteristic length scale $L$, and $\omega$ is the frequency scaled by the wave speed divided by  $L$. An infinite number of identical Dirichlet scatterers, of characteristic dimension $\epsilon L$ and arbitrary cross-sectional shape, are distributed with their centroids, say, at the vertices of a two-dimensional Bravais lattice,
\begin{equation}\label{lattice}
\mathbf{R}=n\mathbf{a}_1+m \mathbf{a}_2,
\end{equation}
where $\mathbf{a}_1$ and $\mathbf{a}_2$ are lattice base vectors and $n$ and $m$ are arbitrary integers. According to Bloch's theorem \cite{Kittel:Book}, waves propagating through the crystal satisfy a Bloch condition in the form
\begin{equation} \label{Bloch}
u(\bx)=U(\bx)e^{i\mathbf{k}\bcdot\bx}, \quad U(\bx+\mathbf{R})=U(\bx),
\end{equation}
where $U$ possesses the periodicity of the lattice and $\mathbf{k}$ is the Bloch wave vector. Given \eqref{Bloch}, we only need to consider an elementary unit cell encapsulating a single scatterer centred at the origin $\bx=0$, to which we attach polar coordinates $(r,\theta)$. The Dirichlet constraint is then written as
\begin{equation}\label{dir}
u = 0 \quad \text{on} \quad r = \epsilon \kappa(\theta),
\end{equation}
where $\kappa(\theta)$ is a shape function. Our goal here is to calculate the dispersion surfaces $\omega=\Omega(\mathbf{k})$. The latter inherit the periodicity of the reciprocal lattice, i.e.~$\Omega(\mathbf{k}+\mathbf{G})=\Omega(\mathbf{k})$, where
\begin{equation}\label{recip}
\mathbf{G}=n\mathbf{b}_1+m\mathbf{b}_2,
\end{equation}
$n$ and $m$ being arbitrary integers, and $\mathbf{b}_1$ and $\mathbf{b}_2$  the reciprocal lattice base vectors defined through $\mathbf{a}_i\bcdot\mathbf{b}_j=2\pi\delta_{ij}$. Additional symmetries further limit the $\mathbf{k}$ vectors required to be considered to the ``irreducible  Brillouin zone'' \cite{Joannopoulos:Book}, which will shall define on a case by case basis.

\subsection{Matched asymptotics}\label{ssec:matching}
Henceforth we consider the asymptotic limit $\epsilon\to0$, with $\omega$ and $\kappa(\theta)=O(1)$. To this end, we conceptually decompose the elementary cell into an ``outer'' domain where $r=O(1)$ and an ``inner'' domain where $\rho = r/\epsilon=O(1)$. As prompted in the introduction, logarithmic terms are treated in par with $O(1)$ terms. Since Bloch eigenmodes are determined only up to a multiplicative constant, we may assume without loss of generality that $u=O(1)$ in the outer domain. The inner region $u$ is at most of comparable magnitude.

Accordingly, we assume an inner expansion of the form
\begin{equation}\label{inner}
u\sim \Phi(\rho,\theta;\epsilon) + a.e.(\epsilon), \quad \rho,\Phi = O(1),
\end{equation}
where ``a.e.'' denotes algebraic error. Substituting \eqref{inner} into \eqref{helm} and \eqref{dir}, we find that $\Phi$ satisfies Laplace's equation
\begin{equation}\label{inner laplace}
\nabla_{\rho}^2\Phi=0,\quad \nabla_{\rho}^2=\frac{1}{\rho}\pd{}{\rho}\left(\rho\pd{}{\rho}\right)+\frac{1}{\rho^2}\pd{^2}{\theta^2},
\end{equation}
the Dirichlet condition 
\begin{equation} \label{inner der}
\Phi=0 \quad \text{on} \quad \rho=\kappa(\theta),
\end{equation}
and matching with the outer region as $\rho\to\infty$. Regardless of the shape function $\kappa(\theta)$, at large distances \cite{Batchelor:Book}
\begin{equation}\label{Phi large}
\Phi \sim A \ln \rho + B + O(1/\rho) \quad \text{as} \quad \rho\to\infty,
\end{equation}
where terms algebraically growing with $\rho$ are disallowed as they become algebraically large in $\epsilon$ in the outer region. In \eqref{Phi large}, $A$ is a constant that once known determines through the solution of the inner problem the constant $B$ and similarly smaller terms in the far-field expansion. For example, for a circular cylinder, $\kappa=1$, the solution is simply $\Phi=A\ln \rho$. Solutions for other cross-sectional shapes are obtained by conformal mapping of the solution for a circular cylinder. Specifically, if $z=\rho \exp(i\theta)$ and $w=f(z)$ transform the domain $\rho>\kappa(\theta)$ to the unit circle, with  $f(z)\sim S z+O(1)$ as $|z|\gg1$ ($S>1$), then $\Phi/A=\mathrm{Re}[\ln w]=\mathrm{Re}[\ln f(z)]$ and hence $\Phi/A\sim \ln |z| + \ln S+O(1/|z|)$ as $|z|\to\infty$. The result of this sidetrack is that, for a general shape, 
\begin{equation}\label{Phi large 2}
\Phi \sim A\ln (S\rho) + O(1/\rho) \quad \text{as} \quad \rho\to\infty,
\end{equation}
where $A$ is a constant to be determined and $S$ is a shape parameter, which has been derived for various cross-sectional shapes in the literature \cite{Hinch:91,Ward:93,Krynkin:09}. When matching with the outer region, only the large $\rho$ expansion \eqref{Phi large 2} is relevant, hence dependence on shape can be alternatively captured by assuming a circular cylinder with effective radius $(\epsilon/S)L$. For simplicity we may therefore set $S=1$, henceforth interpreting $\epsilon$ as the effective radius.

In the outer region we assume the expansion 
\begin{equation}
u\sim \phi(\bx;\epsilon)+ a.e.(\epsilon), \quad r,\phi=O(1)
\end{equation}
where $\phi$ satisfies \eqref{helm}, the Bloch condition \eqref{Bloch}, and matching with the inner region as $r\to0$. The solution is necessarily of the form 
\begin{equation} \label{phi decomp}
\phi=\psi(\bx)+aH_0\ub{1}(\omega r),
\end{equation}
where $\psi(\bx)$ is a solution of \eqref{helm} that is \emph{regular} at $r=0$, $a$ is an undetermined constant, and $H_0\ub{1}$ is the zeroth-order Hankel function of the first kind. The latter is the fundamental outward-radiating solution of \eqref{helm}, which is logarithmically singular as $r\to0$. Higher-order singularities, i.e.~products of constant tensors and gradients of $H\ub{1}_0$ are disallowed at this order since such solutions are algebraically  singular as $r\to0$ and hence become algebraically large in the inner region. 

We may now match the inner and outer expansions using Van Dyke's matching rule (specifically, see pg.~220 in \cite{Van:Book} and \cite{Crighton:73}), treating logarithmic terms, such as $\ln \epsilon$, on par with $O(1)$ terms. Thus, on the one hand, the inner expansion to $O(1)$, written in terms of the outer variable, and then expanded to $O(1)$, is 
\begin{equation}\label{inout}
A \ln (r/\epsilon).
\end{equation}
On the other hand --- noting the regularity of $\psi$ and that
\begin{equation}\label{H0 small r}
H_0\ub{1}(\omega r)\sim \frac{2i}{\pi}\left[\ln (r\omega)+\gamma-\ln 2\right]+1\quad \text{as} \quad r\to0,
\end{equation} 
$\gamma\approx 0.5772$ being the Euler constant --- the outer expansion to $O(1)$, written in terms of the inner variable, and then expanded to $O(1)$, is 
\begin{equation}\label{outin}
\psi(\boldsymbol{0}) + a \left\{\frac{2i}{\pi}\left[\ln (r\omega)+\gamma-\ln 2\right]+1\right\},
\end{equation}
upon rewriting in terms of $r$.
Comparing expansions \eqref{inout} and \eqref{outin}, one finds
\begin{equation}\label{matching a}
a = -\frac{\pi i }{2}\frac{\psi(\boldsymbol{0})}{\ln\frac{2}{\omega\epsilon}-\gamma+\frac{\pi i}{2}}
\end{equation}
and $A = (2i/\pi)a$. Hence, the presence of a scatterer in general causes a perturbation which is $O(1/\ln\frac{1}{\epsilon})$, formally small, but appropriately regarded as being $O(1)$ for all practical purposes as well as  when using the matching rule. 

\subsection{Effective eigenvalue problem}
\label{ssec:effective}
The preceding analysis asymptotically coarse grains the fine details in the vicinity of the scatterer. The resulting effective eigenvalue problem can be formulated in two equivalent ways. Thus, we may seek a regular eigenfunction $\psi$ that satisfies an inhomogeneous Bloch condition owing to the aperiodicity of $aH_0\ub{1}(\omega r)$, $a$ being proportional to $\psi(\boldsymbol{0})$ through \eqref{matching a}. An alternative formulation is obtained by noting from \eqref{phi decomp} that $\phi$ satisfies the forced Helmholtz equation
\begin{equation}\label{phi eq}
\nabla^2\phi +\omega^2\phi = 4ai\delta(\bx),
\end{equation}
where $\delta(\bx)$ denotes the Dirac delta function. In addition, $\phi$ satisfies the Bloch condition \eqref{Bloch} and the asymptotic constraint
\begin{equation}\label{dispersion original}
\lim_{r\to0}\left[\phi(\bx,\left\{\omega,\boldsymbol{k}\right\})+a\frac{2i}{\pi}\ln\frac{\epsilon}{r}\right]=0,
\end{equation}
that follows from the matching condition \eqref{matching a} in conjunction with \eqref{phi decomp}.

\subsection{Empty-lattice Bloch waves and weakly perturbed modes}
\label{ssec:weak}
Preliminary to solving the effective Bloch eigenvalue problem formulated above, we note that \emph{in the absence of scatterers} there are plane-wave solutions of \eqref{helm} that satisfy  the Bloch condition \eqref{Bloch}. Indeed, substituting the generalised Fourier series
\begin{equation}\label{empty u Fourier}
u = \sum_{\mathbf{G}}\mathcal{U}_{\mathbf{G}}\exp\left[i(\mathbf{k}+\mathbf{G})\bcdot\bx\right]
\end{equation}
into \eqref{helm}, where $\mathcal{U}_{\mathbf{G}}$ are constants, we find the empty-lattice dispersion relation
\begin{equation}\label{empty disp}
\omega^2=|\mathbf{k}+\mathbf{G}|^2. 
\end{equation}
It is evident from \eqref{recip} that for any given $\mathbf{k}$ there are infinite number of ``eigenpairs'' $(\omega,\mathbf{k})$. A distinction can be made between those pairs that are simple and those that are degenerate. Generally, we write the empty-lattice eigenmode space as 
\begin{equation}\label{empty u}
u = \sum_{j=1}^D\mathcal{U}_{j}\exp\left[i(\mathbf{k}+\mathbf{G}_j)\bcdot\bx\right],
\end{equation}
where $D$ denotes the level of degeneracy, and $\{\mathbf{G}_j\}_{j=1}^D$ those reciprocal lattice vectors satisfying \eqref{empty disp} for the eigenpair $(\omega,\mathbf{k})$. Since \eqref{empty u} vanishes at the origin if 
\begin{equation}\label{empty one}
\sum_{j=1}^D\mathcal{U}_{j}=0,
\end{equation}
for degenerate eigenpairs we may form a subspace, of dimension $D-1$, of empty-lattice plane-wave solutions that vanish at  $\bx=0$.  Empty-lattice waves in the latter space also satisfy the true eigenvalue problem, with $a=0$ (and hence $\phi=\psi$), and hence are perturbed only weakly --- algebraically in $\epsilon$ --- from the empty-lattice dispersion relation \eqref{empty disp}.  In appendix \ref{app:solvability} it is verified that there are no $\phi$ eigensolutions satisfying \eqref{empty disp} for which $a\ne0$. 

\subsection{Strongly perturbed modes}
\label{ssec:strong}
Consider next the possibility of eigenpairs of the effective eigenvalue problem that do not satisfy the empty-lattice dispersion relation \eqref{empty disp}. Since eigensolutions with $a=0$ necessarily satisfy \eqref{empty disp}, we may assume here that $a\ne0$. 

For the outer field $\phi$ we write the generalised Fourier-series solution
\begin{equation}\label{fourier}
\phi(\bx)/a=\sum_{\mathbf{G}}\phi'_{\mathbf{G}}\exp\left[i\left(\mathbf{k}+\mathbf{G}\right)\bcdot\bx\right],
\end{equation}
which satisfies the Bloch condition \eqref{Bloch}. The coefficients $\phi'_{\mathbf{G}}$ are determined by substituting \eqref{fourier} into \eqref{phi eq} and applying orthogonality relations. Denoting the area of the unit cell by $\mathcal{A}$, we find
\begin{equation}\label{Fourier}
\phi(\bx)/a=\frac{4i}{\mathcal{A}}\sum_{\mathbf{G}}\frac{\exp\left[i\left(\mathbf{k}+\mathbf{G}\right)\bcdot\bx\right]}{\omega^2-|\mathbf{k}+\mathbf{G}|^2}.
\end{equation}
The series solution \eqref{Fourier} is conditionally convergent for $|\bx|\ne\mathbf{0}$ and diverges for $\bx=0$, as would be expected from the asymptotic constraint \eqref{dispersion original}. In appendix \ref{app:doublesum}, we derive the singular asymptotics of \eqref{Fourier},
\begin{equation}\label{asymptotics}
\phi(\bx)/(ai)\sim
\frac{2}{\pi}\left[\ln \frac{r}{2} + \gamma\right]+\sigma(\omega,\mathbf{k})+ o(1) \quad \text{as} \quad r\to0,
\end{equation}
where the limit
\begin{equation}\label{sigma}
\sigma(\omega,\mathbf{k})=\frac{4}{\mathcal{A}}\lim_{R\to\infty}\left[\sum_{|\mathbf{G}|<R}\frac{1}{\omega^2-|\mathbf{k}+\mathbf{G}|^2}+\frac{\mathcal{A}}{2\pi}\ln R\right]
\end{equation}
converges rapidly. Substituting \eqref{asymptotics} into \eqref{dispersion original} furnishes the asymptotic dispersion relation governing the strongly perturbed modes as
\begin{equation}\label{disp}
\sigma(\omega,\boldsymbol{k})+\frac{2}{\pi}\left(\ln\frac{\epsilon}{2}+\gamma\right)=0.
\end{equation}

\subsection{Square and hexagonal lattices}
\label{ssec:curves}
\begin{figure}[t]
   \centering
	\includegraphics[scale=0.33]{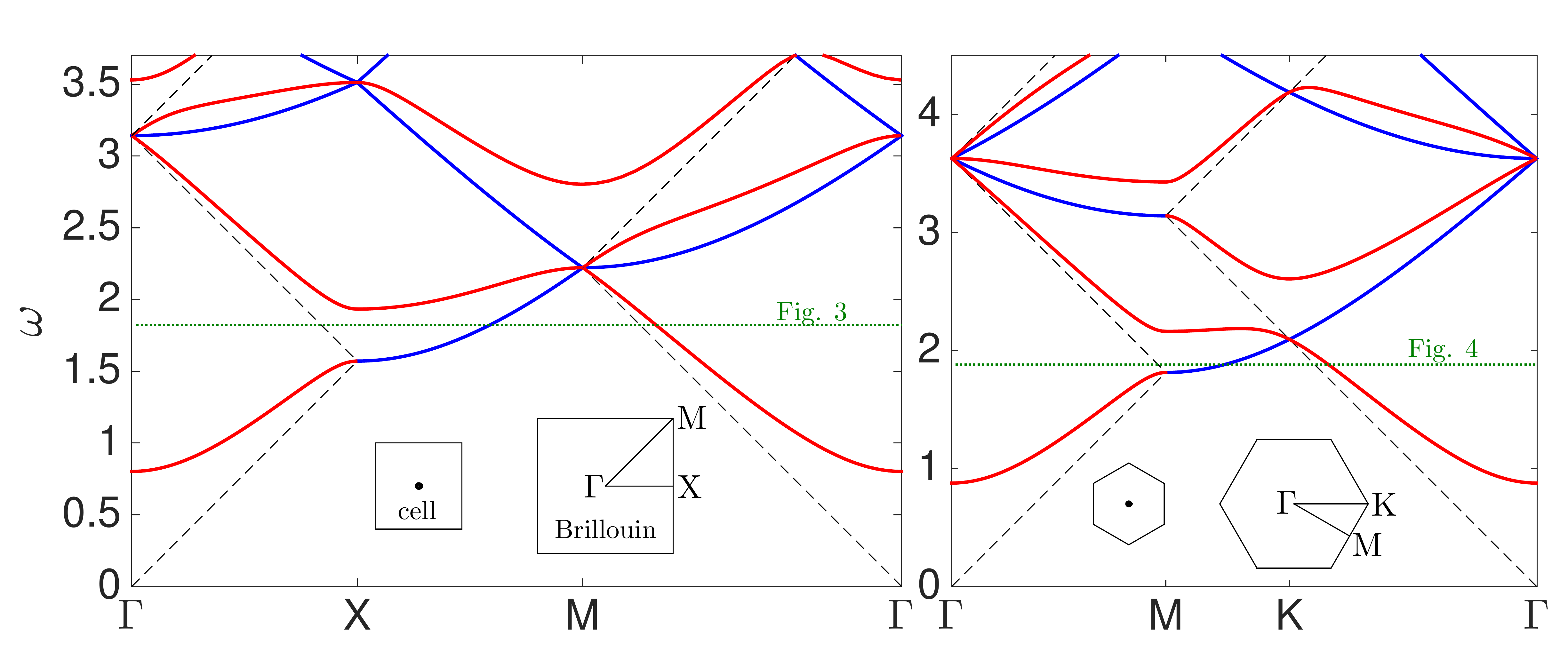} 
   \caption{Dispersion curves for square (left) and hexagonal (right) lattices of Dirichlet scatterers of effective radius $\epsilon=0.05$. Blue and red lines respectively depict weakly and strongly perturbed eigenpairs, the latter coinciding with degenerate empty-lattice eigenpairs. Dotted-black lines depict simple empty-lattice eigenpairs. Inset show unit cells and first Brillouin zone (square --- scale 3:10, hexagonal --- 2:10).}
   \label{fig:disp_sq_hex}
\end{figure}

We now demonstrate the preceding results by calculating the dispersion surfaces for square and hexagonal (or triangular) lattices. Consider first a square lattice,
\begin{equation}\label{square}
\mathbf{a}_1=2\be_x, \quad \mathbf{a}_2=2\be_y, \quad \mathbf{b}_1=\pi\be_x, \quad \mathbf{b}_2=\pi\be_y,  \quad \mathcal{A}=4.
\end{equation}
Following common practice, we plot in Fig.~\ref{fig:disp_sq_hex} (left panel) the dispersion surfaces for $\epsilon=0.05$ along the edges of the irreducible Brillouin zone, which in the present case is bounded by straight lines in reciprocal space connecting the symmetry points 
\begin{equation}
\Gamma=\mathbf{0}, \quad \mathrm{X}=\mathbf{b}_1/2, \quad \mathrm{M}=(\mathbf{b}_1+\mathbf{b}_2)/2.
\end{equation} 
Red solid lines depict strongly perturbed eigenvalues, which are solutions of the dispersion relation \eqref{disp}. Blue solid lines depict weakly perturbed  modes. As shown in \S\S\ref{ssec:weak}, the eigenpairs of the latter are given to algebraic order by the degenerate empty-lattice eigenpairs; they are readily identified from \eqref{empty disp} in conjunction with \eqref{square}. As shown in \S\S\ref{ssec:weak}, if the degeneracy in the absence of a scatterer was $D$, the degeneracy is reduced to $D-1$. The black dashed lines depict simple empty-lattice eigenpairs that are no longer part of the dispersion surfaces.

The distinction between weakly and strongly perturbed empty-lattice waves facilitates interpreting the dispersion diagram. To begin with, the notable zero-frequency band gap is not specific to a square lattice, but in fact exists for any lattice of the class considered in this study. This is because the zero-frequency light line, obtained by substituting $\mathbf{G}=0$ in \eqref{empty disp}, is always simple, hence according to the discussion in \S\S\ref{ssec:weak} the perturbation from it must be appreciable. Next we note that in the empty-lattice case the first $\mathrm{X}$ point, $(\omega,\mathbf{k})=(\pi/2,\mathrm{X})$, is doubly degenerate. In the present case that eigenpair is therefore simple, which explains the partial gap opening above $\mathrm{X}$, with the originally degenerate first band from $\mathrm{X}$ to $\mathrm{M}$ splitting into a weakly perturbed blue curve and a strongly perturbed red curve that joins the second strongly perturbed red curve from $\Gamma$ to $\mathrm{X}$. Also of interest are the high symmetry crossing points at $\mathbf{k}=\Gamma,\mathrm{X}$, and $\mathrm{M}$, where the original four-degeneracy in the empty-lattice case is reduced to three. Dispersion surfaces in the vicinity of degenerate symmetry points are either conical or paraboloidal, while those in the vicinity of simple symmetry points are necessarily paraboloidal \cite{Kittel:Book,Craster:10}. Thus, by noting which surviving empty-lattice curves have zero or conversely nonzero slopes at symmetry points, it becomes possible to qualitatively sketch dispersion diagrams without any computation. 

Clearly, the above qualitative features of the dispersion surfaces are independent of the effective radius $\epsilon$, at least as long as $\epsilon\ll1$. Indeed, it is clear from \eqref{disp} that $\epsilon$ merely affects the magnitude of the strongly perturbed modes.  The dominance of the blue weakly perturbed curves in Fig.~\ref{fig:disp_sq_hex} is misleading; within the irreducible Brillouin zone, rather than along its edges, degenerate empty-lattice eigenpairs are rare. This is clarified by the isofrequency contour shown in Fig.~\ref{fig:anis_sq} for the frequency marked by the horizontal dashed green line in Fig.~\ref{fig:disp_sq_hex}.  

Consider next the triangular (or hexagonal) lattice, with lattice base vectors
\begin{equation}\label{hex a}
\mathbf{a}_1=\be_x-\sqrt{3}\be_y, \quad \mathbf{a}_2=\be_x+\sqrt{3}\be_y, 
\end{equation}
and reciprocal-lattice base vectors
\begin{equation} \label{hex b}
\mathbf{b}_1=\pi\left(\be_x-\frac{1}{\sqrt{3}}\be_y\right), \quad \mathbf{b}_2=\pi\left(\be_x+\frac{1}{\sqrt{3}}\be_y\right).
\end{equation}
The cell area is $\mathcal{A}=2\sqrt{3}$.
The irreducible Brillouin zone is formed by straight lines in reciprocal space connecting the symmetry points 
\begin{equation}\label{hex sym}
{\Gamma}=\mathbf{0}, \quad \mathrm{M}=\mathbf{b}_1/2, \quad \mathrm{K}=|\mathbf{b}_1|/\sqrt{3}.
\end{equation}
Fig.~\ref{fig:disp_sq_hex} (right panel) shows the dispersion curves for $\epsilon=0.05$. Note that the dimension of the eigenmode space at the second $\Gamma$ point is now $5$, having been reduced by one from the empty-lattice degeneracy there, $D=6$. Note also the crossing at the first $\mathrm{K}$ point, where the level of degeneracy has been reduced from $D=3$ to $2$. The latter implies that any additional constraint further reducing the degeneracy there will open an omnidirectional  gap. 

\section{Unit cells occupied by multiple scatterers} \label{sec:multiple}
\subsection{Effective eigenvalue problem}
We here generalise to the case where in each cell there are $P>1$ particles at $\bx=\bx_j$, $j=1\ldots P$. It is assumed that $|\bx_m-\bx_n|\gg \epsilon$ for $m\ne n$; otherwise, the neighbouring scatterers share a common inner region, in which case  we are back to a Bravais lattice with the particle multiplicity captured by an effective $\epsilon$. The analysis of the present case closely follows that of \S\ref{sec:main_analysis}. Thus, the outer potential \eqref{phi decomp} generalises to 
\begin{equation}\label{outer multi}
\phi=\psi(\bx) + \sum_{p=1}^Pa_pH_0\ub{1}(\omega|\bx-\bx_p|),
\end{equation}
where $\psi(\bx)$ is a regular function, and $\phi$ satisfies the Bloch condition \eqref{Bloch}. {It} follows from \eqref{outer multi} that $\phi$ satisfies the governing equation 
\begin{equation}\label{multi helm}
\nabla^2\phi+\omega^2\phi = -\frac{4}{i}\sum_{p=1}^Pa_p\delta(\bx-\bx_p),
\end{equation}
where matching with the inner region of each scatterer provides $P$ conditions, 
\begin{equation}\label{matching multi}
\left[\frac{2i}{\pi}\left(\ln \frac{2}{\omega \epsilon}-\gamma\right)-1\right]a_j=\psi(\bx_j)+\sum_{p\ne j}^Pa_pH_0\ub{1}(\omega |\bx_p-\bx_j|), \quad j = 1\ldots P.
\end{equation}
From \eqref{outer multi}, the right hand side of \eqref{matching multi} is the limit
\begin{equation}
\lim_{\bx\to\bx_j}\left[\phi(\bx)-a_j H_0\ub{1}(\omega|\bx-\bx_j|)\right],
\end{equation}
whereby, using \eqref{H0 small r},  conditions \eqref{matching multi}  simplify to 
\begin{equation}\label{multi matching}
\lim_{\bx\to\bx_j}\left[\phi(\bx)-a_j \frac{2i}{\pi}\ln\frac{\epsilon}{|\bx-\bx_j|}\right]=0, \quad j = 1\ldots P.
\end{equation}

In the case of multiply occupied unit cells, empty-lattice waves survive as weakly perturbed modes mainly at high symmetry points, though in some \emph{incidental} cases also along edges of the irreducible Brillouin zone. We next focus on the strongly perturbed modes, identifying and analysing in \S\S\ref{ssec:honey} the weakly perturbed empty-lattice waves in the context of an important example. 

\subsection{Dispersion relation}
Generalising the Fourier-series solution \eqref{Fourier} to satisfy \eqref{multi helm} gives
\begin{equation}\label{multi Fourier}
\frac{\mathcal{A}}{4i}\phi=\sum_{p=1}^P a_p \exp\left[{i\mathbf{k}\bcdot(\bx-\bx_p)}\right]\sum_{\mathbf{G}}\frac{\exp\left[i\mathbf{G}\bcdot(\bx-\bx_p)\right]}{\omega^2-|\mathbf{k}+\mathbf{G}|^2}.
\end{equation}
The singular asymptotics of the double sum in \eqref{multi Fourier} is obtained by generalising the derivation in appendix \S\ref{app:doublesum}, which gives [cf.~\eqref{asymptotics}]
\begin{multline}\label{multi asymptotics}
\phi/i \sim \left[\frac{2}{\pi}\left(\ln\frac{|\bx-\bx_j|}{2}+\gamma\right)+\sigma(\omega,\mathbf{k})\right]a_j 
\\ + \frac{4}{\mathcal{A}}\sum_{p\ne j}^Pa_p\sum_{\mathbf{G}}\frac{\exp\left[i(\mathbf{G}+\mathbf{k})\bcdot(\bx_j-\bx_p)\right]}{\omega^2-|\mathbf{k}+\mathbf{G}|^2} \quad \text{as} \quad \bx\to\bx_j, \quad j=1\ldots P.
\end{multline}
Substituting \eqref{multi asymptotics} into \eqref{multi matching}, we find a set of $P$ equations for the scattering coefficients $\{a_j\}_{j=1}^P$:
\begin{multline}
\left[\frac{2}{\pi}\left(\ln\frac{\epsilon}{2}+\gamma\right)+\sigma(\omega,\mathbf{k})\right]a_j 
\\ + \frac{4}{\mathcal{A}}\sum_{p\ne j}^Pa_p\sum_{\mathbf{G}}\frac{\exp\left[i(\mathbf{G}+\mathbf{k})\bcdot(\bx_j-\bx_p)\right]}{\omega^2-|\mathbf{k}+\mathbf{G}|^2}=0, \quad j=1\ldots P.
\end{multline}
The dispersion relation is obtained by requiring the determinant of the coefficient matrix to vanish. Clearly, for $P=1$ we retrieve \eqref{disp}. For $P=2$, the dispersion relation can be written as
\begin{equation}\label{dispersion two particles}
\left[\frac{2}{\pi}\left(\ln\frac{\epsilon}{2}+\gamma\right)+\sigma(\omega,\mathbf{k})\right]^2 
= \frac{16}{\mathcal{A}^2}\left|\sum_{\mathbf{G}}\frac{\exp\left[i\mathbf{G}\bcdot(\bx_1-\bx_2)\right]}{\omega^2-|\mathbf{k}+\mathbf{G}|^2}\right|^2.
\end{equation}
The sum on the right hand side is conditionally convergent, yet at a fast algebraic rate. The convergence deteriorates as the two scatterers approach; as already noted, however, the extreme case where the centroid-to-centroid separation is $O(\epsilon)$ is actually covered by the analysis of \S\ref{sec:main_analysis}.

\subsection{Honeycomb lattice}
\label{ssec:honey}
\begin{figure}[t]
   \centering
	\includegraphics[scale=0.5]{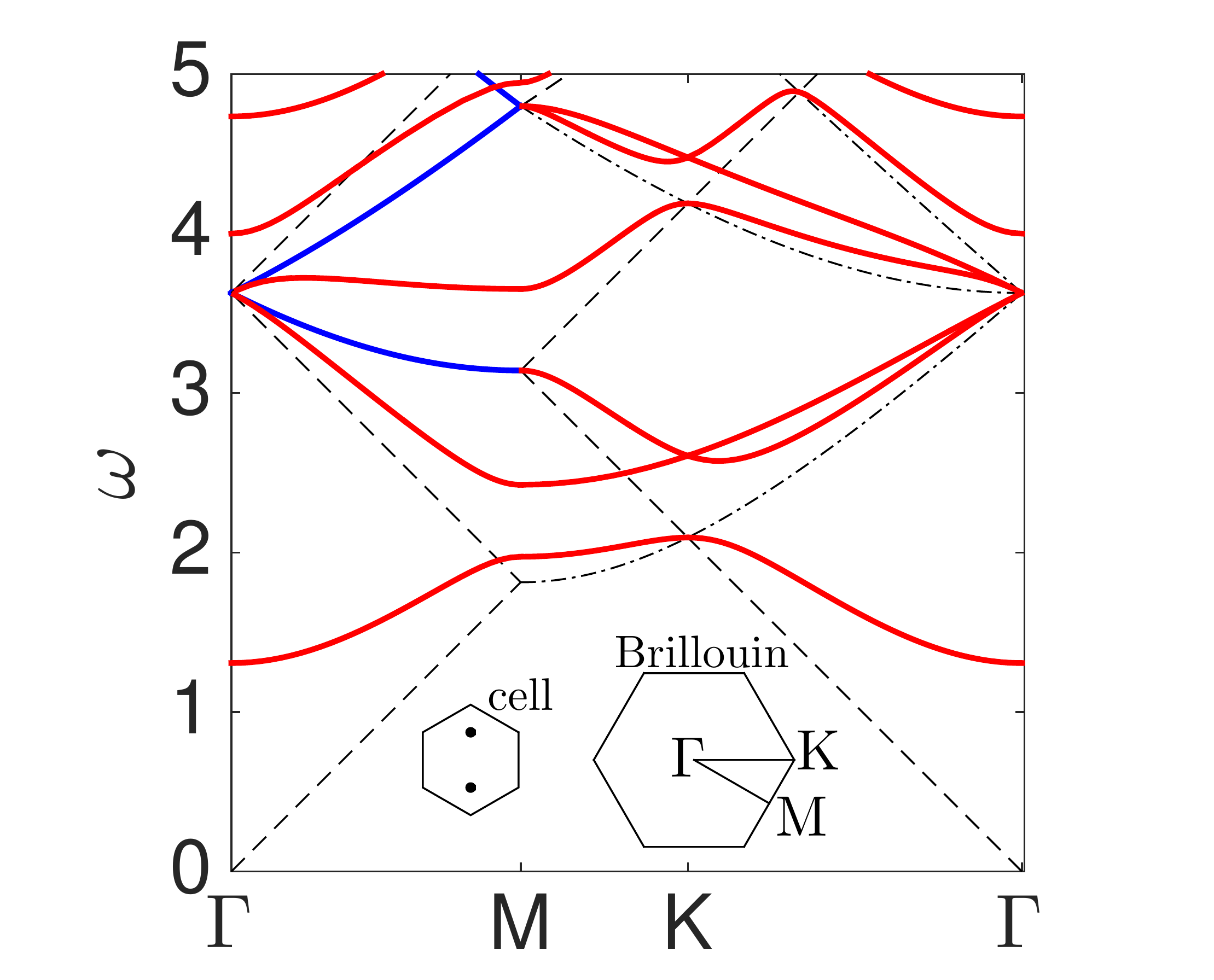} 
   \caption{Same as Fig.~\ref{fig:disp_sq_hex} but for a honeycomb lattice (see \S\S\ref{ssec:honey}). Note that now there are degenerate empty-lattice eigenpairs that do not remain as weakly perturbed modes; the latter are depicted by black dash-dot lines.}
      \label{fig:disp_honey}
\end{figure}

A honeycomb lattice is in fact a triangular lattice with two scatterers in each elementary cell, positioned so that closest neighbours are equidistant. Choosing the elementary cell as the parallelogram generated by the hexagonal base vectors $\mathbf{a}_1$ and $\mathbf{a}_2$ [cf.~\eqref{hex a}], the two scatterers are positioned at
\begin{equation}
\bx_1 = \frac{2}{3}\mathbf{a}_1+\frac{1}{3}\mathbf{a}_2, \quad 
\bx_2 = \frac{1}{3}\mathbf{a}_1+\frac{2}{3}\mathbf{a}_2.
\end{equation}
The dispersion curves are plotted in Fig.~\ref{fig:disp_honey} along the edges of the irreducible Brillouin zone as defined for the underlying triangular lattice in \S\S\ref{ssec:curves}. (The inset shows an alternative choice of the elementary cell.) Red curves depict strongly perturbed modes calculated from the asymptotic dispersion relation \eqref{dispersion two particles}. Blue lines depict empty-lattice eigenpairs that remain as weakly perturbed empty-lattice waves; in contrast, the dotted and dash-dotted lines depict, respectively, simple and doubly degenerate empty-lattice eigenpairs that do not satisfy the asymptotic Bloch problem.

Following the discussion in \S\S\ref{ssec:weak}, one might expect that for doubly occupied unit cells the degeneracy $D$ of empty-lattice eigenpairs reduces to $D-2$. Namely, that no simple or doubly degenerate pairs remain, triply degenerate pairs survive as simple eigenpairs, etc\ldots~ This is consistent with the narrow gap opening up above the first band, the level of degeneracy at the second $\Gamma$ point being reduced to $4$, and the fact that most of the blue curves in the right panel of Fig.~\ref{fig:disp_sq_hex} are replaced by red ones. However, the above rule of thumb  does not universally hold, as demonstrated by the remaining blue curves in Fig.~\ref{fig:disp_honey}, and by the third $\mathrm{M}$ point, which according to the above rule of thumb should have detached from the crossing of the light lines but remains there nevertheless. 

These and other features of the dispersion diagram are understood by examining the existence of empty lattice plane waves that vanish at both $\bx_1$ and $\bx_2$. Alluding to the general form \eqref{empty u} of empty-lattice wave solutions, the conditions for this are
\begin{equation}\label{empty multi}
\sum_{j=1}^D\mathcal{U}_{j}\exp\left(i\mathbf{G}_j\bcdot\bx_1\right)=0, \quad \sum_{j=1}^D\mathcal{U}_{j}\exp\left(i\mathbf{G}_j\bcdot\bx_2\right)=0, 
\end{equation} 
where $D$ and $\mathbf{G}_j=n_j\mathbf{b}_1+m_j\mathbf{b}_2$ are determined from \eqref{empty disp} for any $(\omega,\mathbf{k})$ pair satisfying the empty-lattice dispersion relation. For $D=1$, there are no nontrivial solutions of \eqref{empty multi}. Consistently with the rule of thumb suggested above, for $D>1$, and if the two equations \eqref{empty multi} are independent, the dimension of the space of solutions reduces to $D-2$. If, however, they happen to be dependent, its dimension is $D-1$. 

As an example, consider for example $\mathbf{k}$ traversing from $\Gamma$ to $\mathrm{M}$. Setting $\mathbf{k}=t\mathbf{b}_1/2$, $0\le t\le1$, the empty-lattice dispersion relation \eqref{empty disp} becomes
\begin{equation}\label{empty GM}
\frac{3\omega^2}{\pi^2}= 4 m^2 + 2 m (2 n + t) + (2 n + t)^2.
\end{equation}
From \eqref{empty GM} we see that the second $\Gamma$ point is degenerate with $D=6$, and $\{(n_j,m_j)\}_{j=1}^6$ $=\{(0,\pm1),(\pm1,0),(1,-1),(-1,1)\}$. For $0<t\le 1$, the third light line is degenerate with $D=2$ and $\{(n_j,m_j)\}_{j=1}^2=\{(0,-1),(-1,1)\}$. In the former case, \eqref{empty multi} read
\begin{gather}
\mathcal{U}_{1}e^{i\frac{2}{3}\pi}+\mathcal{U}_{2}e^{-i\frac{2}{3}\pi}+\mathcal{U}_{3}e^{i\frac{4}{3}\pi}+\mathcal{U}_{4}e^{-i\frac{4}{3}\pi}+\mathcal{U}
_{5}e^{i\frac{2}{3}\pi}+\mathcal{U}_{6}e^{-i\frac{2}{3}\pi}=0, \\
\mathcal{U}_{1}e^{i\frac{4}{3}\pi}+\mathcal{U}_{2}e^{-i\frac{4}{3}\pi}+\mathcal{U}_{3}e^{i\frac{2}{3}\pi}+\mathcal{U}_{4}e^{-i\frac{2}{3}\pi}+\mathcal{U}_{5}e^{-i\frac{2}{3}\pi}+\mathcal{U}_{6}e^{i\frac{2}{3}\pi}=0,
\end{gather}  
which are independent. Thus, as predicted by the rule of thumb, the second $\Gamma$ point is four-degenerate. 
In contrast, in the latter case, \eqref{empty multi} read
\begin{gather}
\mathcal{U}_{1}e^{-i\frac{2}{3}\pi}+\mathcal{U}_{2}e^{-i\frac{2}{3}\pi}=0, \\
\mathcal{U}_{1}e^{-i\frac{4}{3}\pi}+\mathcal{U}_{2}e^{i\frac{2}{3}\pi}=0,
\end{gather}  
which are clearly dependent, explaining the lower blue curve in Fig.~\ref{fig:disp_honey}.

\begin{figure}[t]
   \centering   
	\includegraphics[scale=0.35]{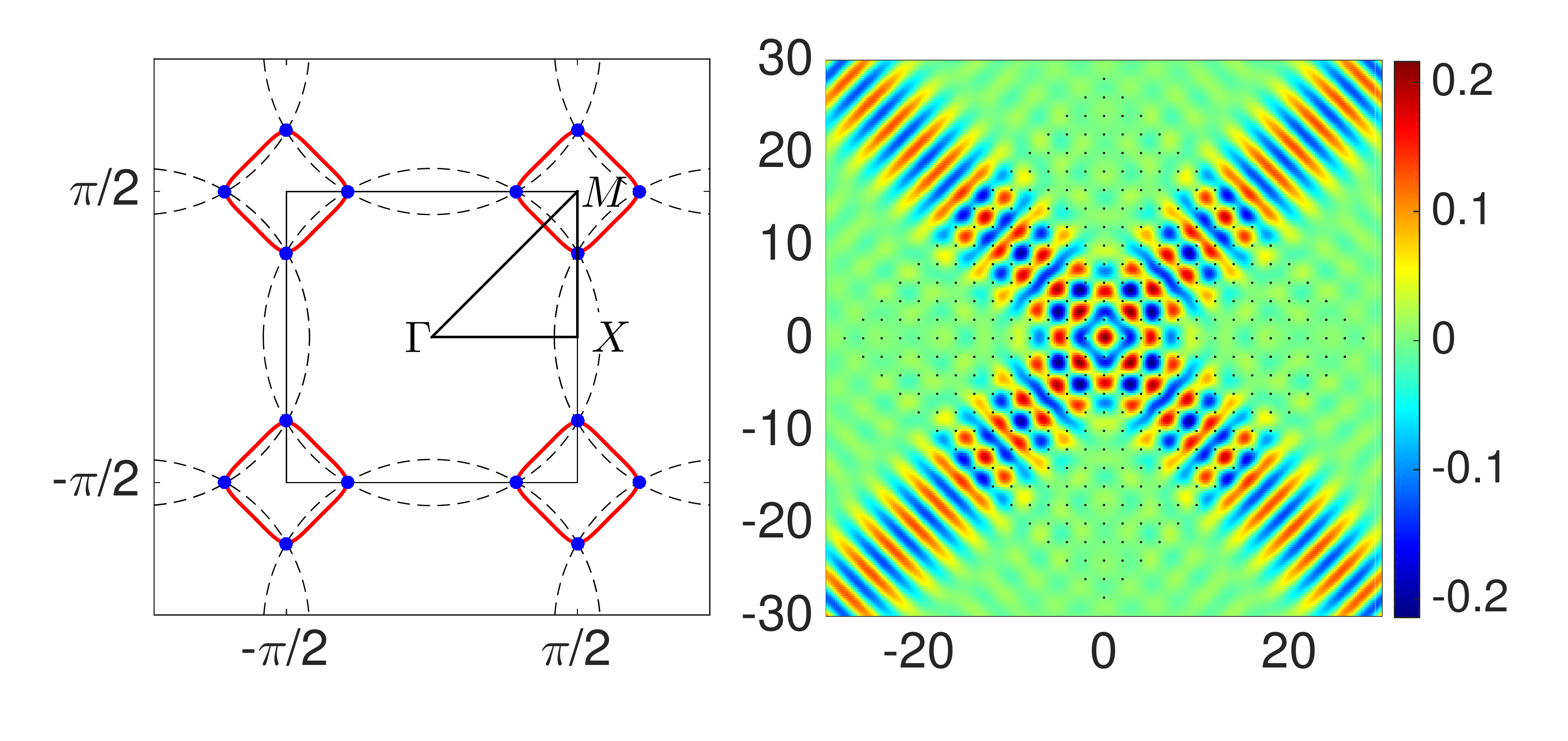}    
\caption{Dynamic anisotropy. Scattering at $\omega=1.82$ from a finite square lattice of Dirichlet inclusions of radius $\epsilon=0.05$, subjected to a point-source incident field at the origin $u_i=H_0\ub{1}(\omega r)$. Left panel: Isofrequency contour of the corresponding infinite lattice (Blue points --- weakly perturbed eigenpairs; Dashed lines --- empty-lattice eigenpairs). Right panel: $\mathrm{Re}[u]$.}
\label{fig:anis_sq}
\end{figure}

\section{Scattering by a finite collection of scatterers} \label{sec:scat}
There is a close connection between the Bloch dispersion problem, which is defined on a unit cell of an infinite lattice, and the  scattering properties of a truncated finite variant of the same lattice. In order to demonstrate this in the context of the dispersion surfaces calculated in the preceding sections, we shall employ a scattering formulation known as Foldy's method \cite{Foldy:45}, which here readily follows from the inner-outer asymptotic procedure of \S\ref{ssec:matching}. To be specific, we consider scattering from $N$ Dirichlet scatterers of effective radius $\epsilon$, positioned at $\bx=\{\bx_l\}_{l=1}^N$, and subjected to an incident field $u_i(\bx)$. The field $u$ satisfies the Helmholtz equation \eqref{helm}, Dirichlet condition $u=0$ on the boundary of each scatterer, and a radiation condition at large distances on the scattering field $u-u_i$. In the limit $\epsilon\ll1$, the ``outer'' solution of the scattering problem is readily seen to be [cf.~\eqref{phi decomp} and \eqref{matching a}]
\begin{equation} \label{phi decomp finite}
u \sim u_i(\bx)+\sum_{l=1}^{N}a_lH_0\ub{1}(\omega r_l) + a.e.(\epsilon),
\end{equation}
where $r_l=|\bx-\bx_l |$ and the coefficients $a_l$ are determined from $N$ matching conditions,
\begin{equation}\label{matching a finite}
a_l + \frac{\pi i }{2}\frac{1}{\ln\frac{2}{\omega\epsilon}-\gamma+\frac{\pi i}{2}}\left[u_i(\bx_l)+\sum_{p\ne l} a_p H_0\ub{1}(\omega|\bx_l-\bx_p|)\right]=0, \quad l=1\ldots N.
\end{equation}
Numerically solving the above linear system is fairly straightforward, even for a large number of scatterers. Once the coefficients $\{a_l\}_{l=1}^N$ have been determined, the field is obtained from \eqref{phi decomp finite}. The above approximate scheme was derived by Foldy, who conjectured isotropic scattering and obtained the ``scattering coefficient'' --- essentially the $\epsilon$-dependent prefactor in \eqref{matching a finite} --- in a semi-heuristic manner. The validity of Foldy's assumption and scattering coefficient was later confirmed by reducing, in the long-wavelength limit, the exact solution for an isolated scatterer \cite{Martin:14}, and also by asymptotic matching \cite{Martin:06}. Foldy's approach is applicable, with appropriate modifications, not only to the Helmholtz--Dirichlet problem, but whenever the material properties, and the type of waves considered, are such that the reaction of each scatterer is dominantly isotropic. For the Biharmonic--Dirichlet problem mentioned in the introduction Foldy's method is particularly intuitive \cite{Evans:07}, since the equations analogous to \eqref{matching a finite} are obtained by applying regular Dirichlet point constraints. 

\begin{figure}[t]
   \centering
        \includegraphics[scale=0.35]{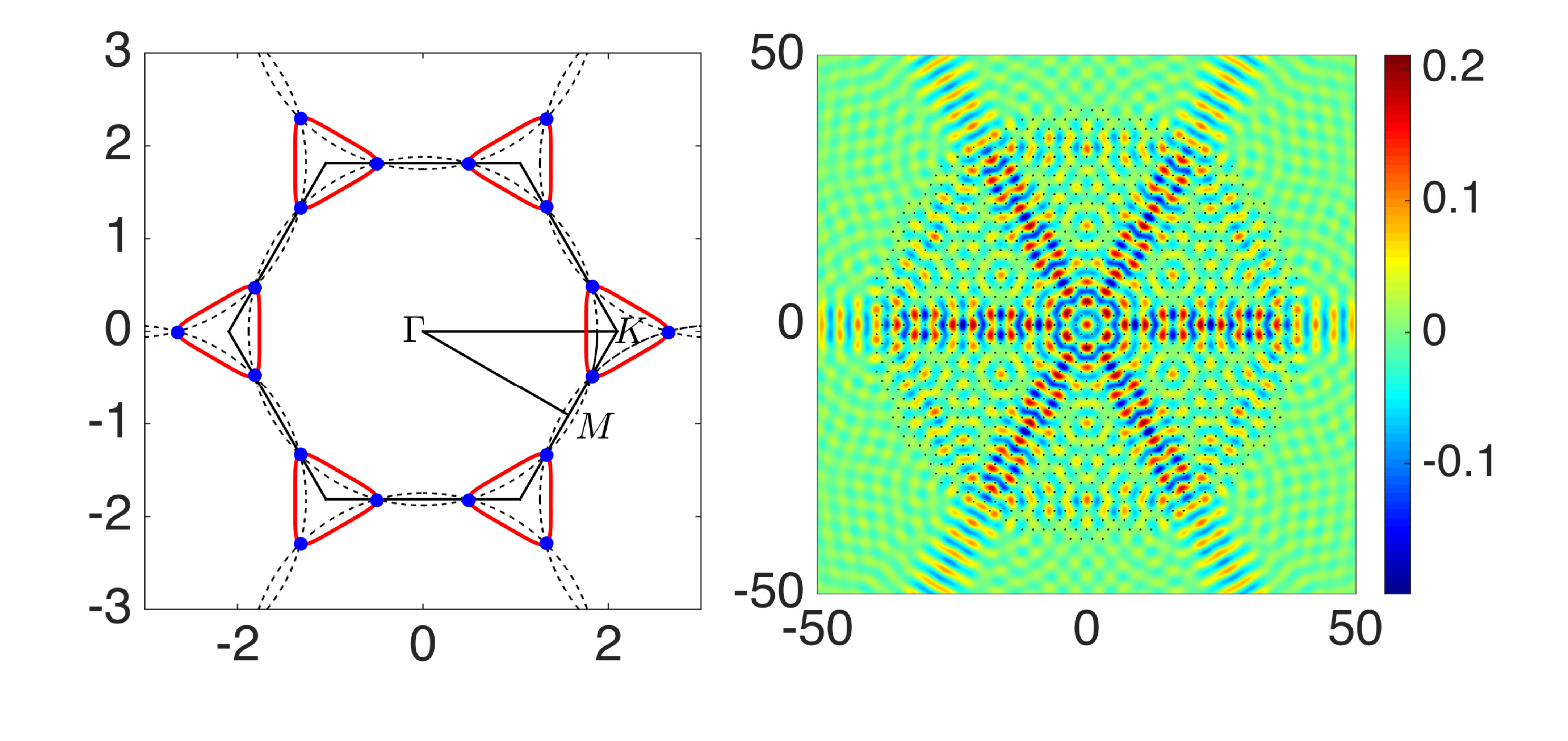}
\caption{Same as in Fig.~\ref{fig:anis_sq} but for a hexagonal lattice at $\omega=1.88$.}
        \label{fig:anis_hex}
\end{figure}

Here we employ Foldy's method to demonstrate the strong dynamic anisotropy implied by the dispersion relations calculated in \S\ref{ssec:curves}. To this end, we solved \eqref{matching a finite} for finite square and hexagonal lattices, with a point source replacing a centrally located scatterer. In Figs.~\ref{fig:anis_sq} and \ref{fig:anis_hex}, the right panels depict the total fields $\mathrm{Re}[u]$, calculated by solving \eqref{matching a finite} with a point source at the origin, $u_i=H_0\ub{1}(\omega r)$, the frequencies being $\omega=1.82$ and $\omega=1.88$ in the square (Fig.~\ref{fig:anis_sq}) hexagonal cases (Fig.~\ref{fig:anis_hex}). The left panels show in reciprocal lattice space the corresponding iso-frequency contours, calculated from the infinite-lattice dispersion relation \eqref{disp}. The latter complement the dispersion diagrams in Fig.~\ref{fig:disp_sq_hex}, and are particularly convenient for interpreting the behaviour of finite lattices. At the selected frequencies, the dispersion contours are approximately straight lines connecting the crossing points of the empty-lattice light circles. The group velocities, which give the permissible directions for energy propagation \cite{Sakoda:Book}, are normal to the isofrequency contours, hence the highly directional response.

\section{Discussion} \label{sec:disc}
In this paper we studied wave propagation in lattices of subwavelength Dirichlet scatterers, the periodicity of the lattice  being on the order of the wavelength. Our method entails: (i) effectively replacing the finite scatterers by {singular point constraints} by applying the method of matched asymptotic expansions; (ii) identifying the space of  empty-lattice waves that are weakly perturbed; and (iii) deriving a dispersion relation for the strongly perturbed modes by extracting the singular ``inner'' asymptotics of a generalised Fourier-series solution of the effective ``outer'' eigenvalue problem. Using this method, we generated dispersion curves for the fundamental square, hexagonal, and honeycomb lattices, pointing out the essential role played by degeneracy and the existence of weakly perturbed modes. We also demonstrated using Foldy's method the strong dynamical anisotropy implied by those dispersion curves. Our method can be readily employed to study more involved lattices of small Dirichlet scatterers. In particular, the topological properties of lattices breaking mirror symmetry are currently under intensive investigation \cite{Lu:16,Ma:16}. In the present formulation this can be easily achieved by asymmetrically positioning two or more scatterers in a unit cell, or, alternatively, assigning symmetrically positioned scatterers different effective radii. 

The two dimensional Dirichlet--Helmholtz problem considered herein, which has realisations in electromagnetics and acoustics, serves to demonstrate that small scatterers are not necessarily weak scatterers (though they are e.g.~in the 3D variant of this problem and for Neumann scatterers in both 2D and 3D). We already mentioned in the introduction the Biharmonic pinned-plate problem, and other important examples include wire media \cite{Pendry:96,Belov:02,Simovski:12}, Faraday cages \cite{Martin:14,Chapman:15,Hewett:16}, lattices of high-contrast rods \cite{Schuller:07,Peng:07}, and plasmonic nanoparticle waveguides and metasurfaces \cite{Maier:07}. In addition, ongoing research into photonic and mechanical analogies of topological insulators and the integer quantum hall effect has stirred interest in media breaking time-reversal symmetry. This has led to novel metamaterial designs incorporating small-scale resonant mechanical components \cite{Torrent:13,Wang:15}, and high-contrast opto-magnetic rods \cite{Wang:08}. We expect that the type of analysis carried out herein can be adopted to study all of these examples, and in general media consisting of wavelength-scale lattices built out of small elements that strongly scatter. For any given example, this would entail repeating the inner-outer asymptotics, which enables replacing the small scale elements by point constraints; this could be much more involved than in the present case \cite{Lindsay:15}, especially if the strong scattering results from a subwavelength resonance. Asymptotic coarse-grained descriptions of such media would not only be technically advantageous, but may also offer new insight by highlighting the existence of both weakly and strongly perturbed modes. 

It is worth noting that the Foldy methodology, although often employed in studies of random scattering \cite{Martin:08,Maurel:10}, appears somewhat under-employed in the periodic photonic, phononic and platonic literature. As demonstrated here it provides an ideal setting, all be it limited to small scatterers, for investigating many of the phenomena of interest (such as for the dynamic anisotropy shown in 
Figs. \ref{fig:anis_sq} \& \ref{fig:anis_hex}) and their dependence upon, say, lattice geometry in an algebraic setting; the solutions are accessed far more rapidly than, say, the more usual finite element approaches \cite{Nicolet:04}, popularised by COMSOL and other commercial packages, commonly used that require refined meshes for very small scatterers.

\appendix
\section{Solvability condition}
\label{app:solvability}
Consider the effective eigenvalue problem formulated in \S\S\ref{ssec:effective} for the outer field $\phi$, in the case where $(\omega,\mathbf{k})$ satisfy the empty-lattice dispersion relation \eqref{empty disp}. The corresponding space of empty-lattice plane waves is given by \eqref{empty u}, and we choose an arbitrary plane wave $u$ from that space. The complex conjugate of the latter, $u^*$, satisfies 
\begin{equation}\label{u conj}
\nabla^2 u^* +\omega^2 u^* =0
\end{equation}
and the Bloch condition \eqref{Bloch} at frequency $\omega$ and Bloch wave vector $-\mathbf{k}$. Subtracting \eqref{u conj} multiplied by $\phi$ from \eqref{phi eq} multiplied by $u^*$, followed by an integration over the unit cell, yields
\begin{equation}
\iint \left(u^*\nabla^2\phi-\phi\nabla^2 u^*\right)\,dA = {4ia}u^*(\mathbf{0}).
\end{equation} 
Using green's theorem, and that $\phi \exp(-i\mathbf{k}\bcdot\bx)$ and $u^*\exp(i\mathbf{k}\bcdot\bx)$ both possess the periodicity of the lattice, the left hand side can be shown to vanish. We therefore find, upon substituting \eqref{empty u}, 
\begin{equation}
a \sum_{j=1}^{D}\mathcal{U}_j = 0.
\end{equation}
Since we may choose the $\mathcal{U}_j$'s arbitrarily, it must be that $a=0$. 

Hence, to conclude, when $(\omega,\mathbf{k})$ satisfy the empty-lattice dispersion relation \eqref{empty disp}, only $\phi$ eigensolutions with $a=0$ --- necessarily empty-lattice plane waves --- are permitted. Furthermore, it then follows from \eqref{dispersion original} that only the empty-lattice plane waves that vanish at the position of the scatterer, as characterised in \S\S\ref{ssec:weak}, remain unperturbed to algebraic order in $\epsilon$. 

\section{Double-sum asymptotics}
\label{app:doublesum}
We here derive the asymptotics of the outer Fourier-series solution \eqref{Fourier} in the limit where $r=|\bx|\to0$. As already noted, substituting $\bx=0$ yields a diverging sum and hence this limit is singular. To overcome this, we separately sum terms corresponding to reciprocal lattice vectors $\mathbf{G}$ whose magnitudes are smaller and larger than an arbitrary large radius $R$ satisfying $1\ll R\ll 1/r$. This gives, after obvious approximations, 
\begin{equation}\label{as 1}
\frac{\mathcal{A}}{4ia}\phi(\bx)\sim
\sum_{|\mathbf{G}|<R}\frac{1}{\omega^2-|\mathbf{k}+\mathbf{G}|^2}
-\sum_{|\mathbf{G}|>R}\frac{\exp\left(i\mathbf{G}\bcdot\bx\right)}{|\mathbf{G}|^2}+o(1) \quad \text{as} \quad r\to0.
\end{equation}
Both sums in \eqref{as 1} diverge logarithmically as $R\to\infty$ but the singularity must cancel out. Writing $\mathbf{G}\bcdot\bx=Gr\cos\theta$, the second sum in \eqref{as 1} becomes 
\begin{equation}
\sum_{|\mathbf{G}|>R}\frac{\exp\left(i\mathbf{G}\bcdot\bx\right)}{|\mathbf{G}|^2}=\sum_{G>R}\frac{\exp\left(iGr\cos\theta\right)}{G^2},
\end{equation}
which is approximated with algebraic error by a Riemann sum,
\begin{equation}
\sim \frac{1}{\mathcal{A}_G}\int_{R}^{\infty}\int_0^{2\pi}\frac{\exp\left(iGr\cos\theta\right)}{G}d\theta dG +o(1),
\end{equation}
where $\mathcal{A}_G=4\pi^2/\mathcal{A}$ denotes the area of a unit cell in reciprocal space. Integrating with respect to $\theta$ gives after a change of variables $\xi=G/R$,
\begin{equation}
\sim \frac{2\pi}{\mathcal{A}_G}\int_1^{\infty}\xi^{-1}J_0(rR\xi)\,d\xi +o(1).
\end{equation}
Standard asymptotic evaluation of the latter integral for $rR\ll1$ gives
\begin{equation}
\sim\frac{2\pi}{\mathcal{A}_G}\left(\ln\frac{2}{rR}-\gamma\right)+o(1).
\end{equation}
Substituting into \eqref{as 1}, and defining the limit \eqref{sigma}, we find the asymptotic result \eqref{asymptotics} stated in the text. The derivation in the case of multiply occupied cells follows along the same lines. 

\bibliographystyle{siamplain}
\bibliography{refs.bib}

\end{document}